\documentclass[%
 reprint,
 amsmath,amssymb,
 aps,
]{revtex4-1}

\usepackage{graphicx}
\usepackage{dcolumn}
\usepackage{bm}
\usepackage{color}
\usepackage[normalem]{ulem}

\graphicspath{{/}{Figs/}}
\begin{document}
\title{Full control of quantum Hall supercurrent in a side gated graphene Josephson junction}

\author{Andrew Seredinski$^1$} \email{ams168@duke.edu}
\author{Anne W. Draelos$^1$}
\author{Ethan G. Arnault$^1$}
\author{Ming-Tso Wei$^1$}
\author{Hengming Li$^2$}
\author{Kenji Watanabe$^3$}
\author{Takashi Taniguchi$^3$}
\author{Fran\c cois Amet$^2$}
\author{Gleb Finkelstein$^1$}
\affiliation{%
 $^1$Department of Physics, Duke University, Durham, NC 27708, USA\\
$^2$Department of Physics and Astronomy, Appalachian State University, Boone, NC 28607, USA\\
$^3$Advanced Materials Laboratory, NIMS, Tsukuba 305-0044, Japan
}%
\date{\today}

\begin{abstract}
We present a study of a graphene-based Josephson junction with dedicated side gates carved from the same sheet of graphene as the junction itself. These side gates are highly efficient, and allow us to modulate carrier density along either edge of the junction in a wide range. In particular, in magnetic fields in the $1-2$ Tesla range, we are able to populate the next Landau level, resulting in Hall plateaus with conductance that differs from the bulk filling factor. When counter-propagating quantum Hall edge states are introduced along either edge, we observe supercurrent localized along that edge of the junction. Here we study these supercurrents as a function of magnetic field and carrier density.

\end{abstract}

\maketitle

The interplay of spin-helical states and superconductivity is predicted to enable access to non-Abelian excitations such as Majorana zero modes (MZM) \cite{Alicea,Fu2,Lutchyn,Oreg}. Through braiding operations which reveal nontrivial exchange statistics, these states may form the basis for quantum computing architectures which take advantage of topological protections to achieve fault-tolerance \cite{Nayak}. Several technologies to this end are in development, including hybrid superconductor-semiconducting nanowire and superconductor-topological insulator structures \cite{Lutchyn2, Aguado}. Interest in topological superconductivity has also spurred a flurry of activity at the interface of superconductivity and the quantum Hall (QH) effect \cite{Amet,Shalom,Draelos,Seredinski,Calado,Wan,Lee,Park,Sahu}. It has been predicted that  quasi-1D superconducting contacts to a QH structure could enable MZM and parafermions \cite{Clarke1,Clarke2,Mong}.

Heterostructures of graphene and hexagonal boron nitride with one-dimensional superconducting contacts \cite{Calado} can demonstrate a remarkable contact transparency, allowing us to observe supercurrent in the quantum Hall regime \cite{Amet}. However, the microscopic details of the supercurrent in the QH regime remain an open subject \cite{Draelos}. In particular, the nature of the superconducting coupling to the edge states could depend e.g. on the vacuum edges of the graphene mesa, the drift velocity of the QH edge states, or the presence of incompressible strips. 
Yet, the electrostatic potential along the mesa edge is typically poorly controlled; it is known to be influenced by charge accumulation effects \cite{Silvestrov}, and may be strongly affected by the disorder resulting from physical etching. Here, we examine a graphene Josephson junction with two side gates that allow us to directly manipulate the QH edge states. By tuning either gate, we can change the Landau level (LL) filling factor along the edges in a wide range. We controllably induce counter-propagating states along either edge and observe supercurrent localized solely along one edge. This technique holds promise for making future devices which would allow one to create and manipulate MZM and parafermions along the lines proposed in \cite{Clarke1,Clarke2}.

	\begin{figure}[]
		\center \label{fig1}
		\includegraphics[width=3.5in]{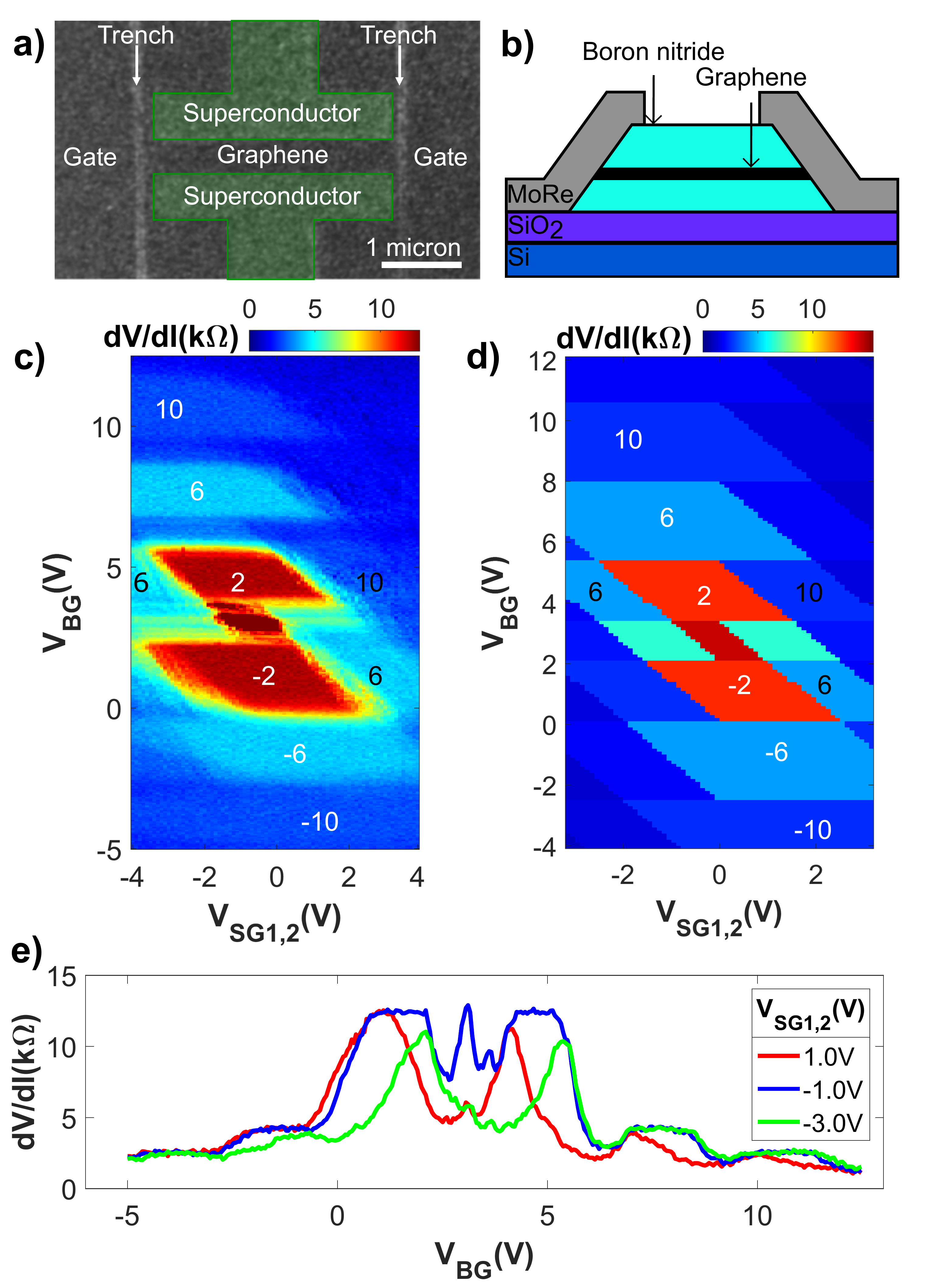}
		\caption {Device layout and gate influence on quantum Hall plateaus. a) SEM micrograph of the device prior to reactive ion etching. MoRe contacts are outlined and colored green for contrast. Two trenches (light gray), $\sim 60$ nm wide, separate the junction from the side gates. The MoRe contacts are spaced from the trenches by $\sim 100$ nm wide regions of graphene, preventing direct contact between MoRe and the edge of the mesa. b) Schematic side-view of a vertical cross-section of (a). c) Resistance map as a function of back gate voltage, $V_{BG}$, and symmetrically applied side gate voltages, $V_{SG1}=V_{SG2}$, at $B=1.8$ T. The diamond-shaped regions correspond to the plateaus of quantized resistance. Their horizontal boundaries (affected by $V_{BG}$ only) correspond to constant electron density in the bulk. The inclined side boundaries of the diamonds correspond to constant filling factors near the edges, where the influences of the back and the side gates compensate each other. The white numbers mark the sample's filling factor, while the black numbers at high sidegate mark sample conductance in units of $e^2/h$. d) Finite element electrostatic simulation of (c) reproducing the diamond shaped regions of constant conductance. The conductance plateaus marked in (c) are marked similarly. Computational details are provided in the supplementary material. e) Sample resistance as a function of $V_{BG}$ at several $V_{SG1,2}$, corresponding to vertical cross-sections of (c). The curves show that the quantum Hall plateaus are best developed with the side gates set to $-1$ V. At $V_{SG1,2}=-3$V and $+1$ V, the plateaus shrink and become asymmetric between the electron and hole-doped sides, as is often found in samples without side gate control.}
	\end{figure}

Our samples are made from graphene encapsulated in hexagonal boron nitride (BN), which protects devices from processing contamination and can yield ballistic transport over micron scales \cite{Mayorov}. The graphene-BN stack is then etched, and quasi-one dimensional contacts to the exposed regions are fabricated~\cite{Dean}. We use molybdenum rhenium (MoRe), a type-II superconductor with an upper critical field of at least 9 T and critical temperature of $\sim 9$ K. The $3$ $\mu$m wide contacts are separated by 500 nm and are initially made to an extended region of graphene. At the next stage, both the junction and the side gates are formed by etching narrow trenches on each side of the contacts (Figure 1a). Applying voltage to the graphene regions that form the side gates allows us to efficiently control the electron density along the edges of the junction. It is important that the etched trenches do not overlap with the contacts, and are instead spaced from them by a graphene strip $\sim$ 100 nm wide. This strip separates the contacts from any atomic-scale spurious states that may exist along the etched edge. 
For consistency, we present results from one Josephson junction; additional measurements of a second device are shown in the supplementary. 

\section{Side gate influence in the quantum Hall regime}

As a magnetic field $B$ is applied perpendicular to the sample, the junction enters the quantum Hall regime. By 1.8 T, the quantum Hall effect is very well developed and we stay at that field in Figures 1-3. The influence of the side gates is significant in this regime, since the edges of the device dominate the transport properties. Figure 1c maps the influence of the back gate and the two side gates, applied symmetrically, $V_{SG1}=V_{SG2}$. This and subsequent measurements in this section are performed with a DC bias current of 10 nA, enough to suppress any supercurrent that may be flowing between the contacts in the QH regime. An additional, negligibly small AC current of 50 pA is applied in order to measure the differential resistance with a lock-in amplifier. The large central red (high resistance) features in Figure 1c mark the $\nu=\pm2$ quantum Hall plateaus. Above and below these are the standard $\nu=\pm6$ states. Only the $\nu\,=\,4(n+\frac{1}{2})$ sequence of filling factors is visible at this field. 

The regions of quantized conductance have a diamond shape, whose boundaries in the back gate direction are flat (horizontal), which means that they are not affected by the side gates. The inclined side boundaries of the red diamonds indicate that they depend both on the side gates and the back gate. These boundaries are interpreted as a line of constant carrier density along the edges of the device, $n_{side} \propto (V_{SG1,2}-\alpha V_{BG}) = const$ where $\alpha \sim 2$ is a constant determined by the relative gate efficiencies. The overall shape of the map in Figure 1c is well reproduced by a simple electrostatic simulation, as shown in Figure 1d.

Finally, the centers of diamond-shaped plateaus in Figure 1c are shifted from $V_{SG1,2}=0$ V indicating that the ``neutral'' side gate voltage is close to -1 V. This differs from the back gate position of the charge neutrality point (3.5 V) not only in magnitude but in polarity, indicating a carrier buildup along the edges of the junction distinct from the doping of the bulk. The side gate influence is illustrated in Figure 1e, which demonstrates that the resistance plateaus of the device, as a function of back gate, are better formed at $V_{SG1,2}=-1$ V than at -3 V or +1 V. 

\begin{figure}[]
		\center \label{fig2}
		\includegraphics[width=3.5in]{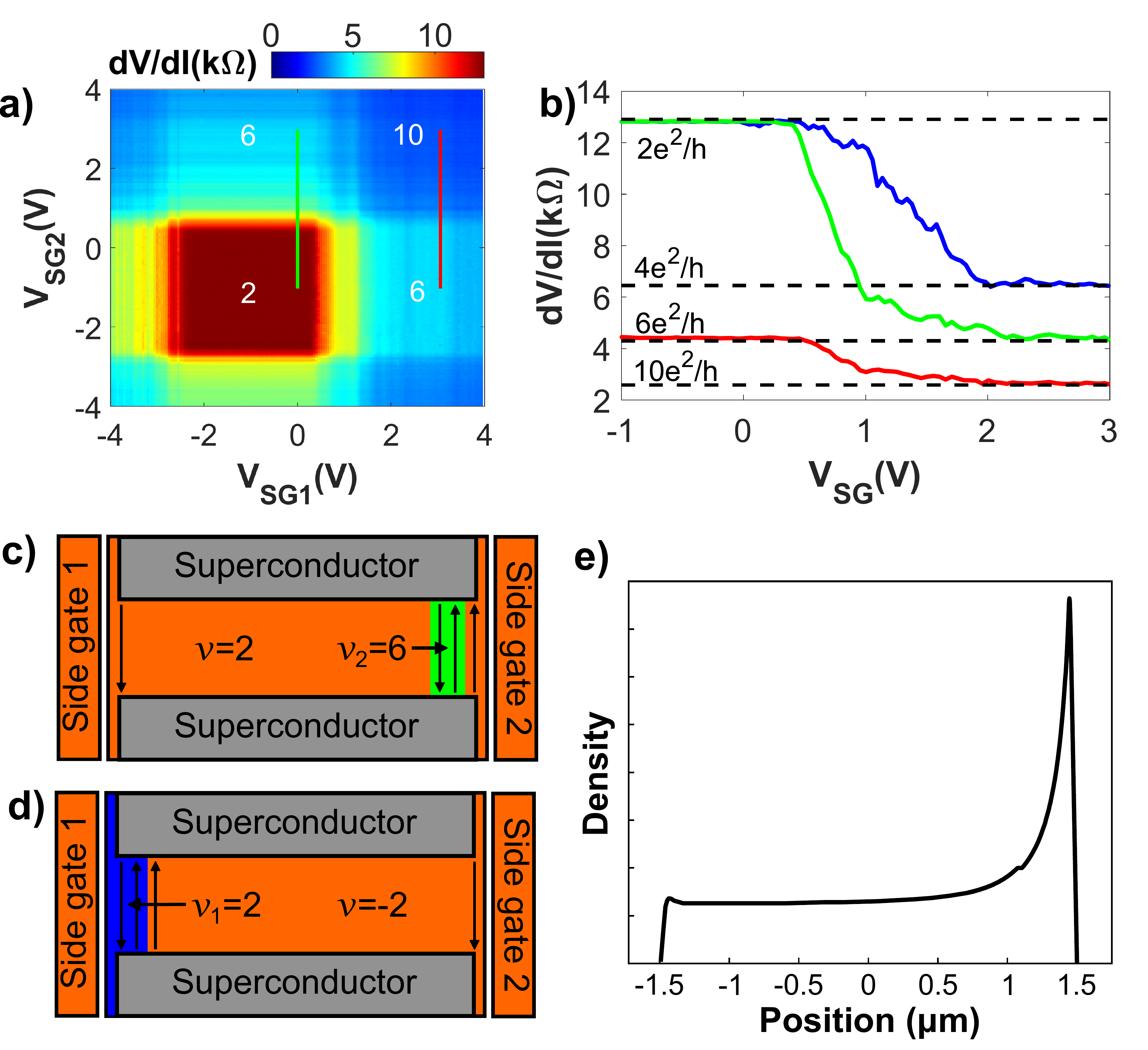}
		\caption{Side gate induced quantum Hall plateaus. a) $dV/dI$ map plotted vs. side gate voltages $V_{SG1}$ and $V_{SG2}$ at $B=1.8$ T. The back gate voltage is fixed at $V_{BG}=4.7$ V, corresponding to the bulk $\nu=2$ state. The numbers mark sample conductance in units of $e^2/h$. b) Sample resistance measured as a function of a single side gate. Green and red curves correspond to the vertical lines in panel (a) at $V_{SG1}= 0$ and 3 V, respectively (with $V_{BG}=4.7$ V). Blue curve shows a similar trace with a bulk filling factor $\nu=-2$ ($V_{BG}=1.5$ V), sweeping $V_{SG1}$ with $V_{SG2}=0$ V. c,d) Schematics corresponding to the green and blue curves in (b) for $V_{SG}$ greater than $\sim 2$ V. Additional edge channels are created near the gate, with local filling factor $\nu_{2}=6$ (c, green region) and $\nu_{1}=2$ (d, blue region). Additional conductance is equal to $4e^2/h$ and $2e^2/h$ in (c) and (d), respectively, on top of the base conductance of $2e^2/h$, as is observed for the blue and green curves in panel (b). e) Schematic of the carrier density within the graphene junction as a function of position when side gate 2 (1) is active (passive), akin to (c).}
	\end{figure}

More insight into the device's phenomenology is gained by applying the side gates independently. Figure 2a shows a resistance map of the device as a function of both side gates at $V_{BG}=$ 4.7 V. (Taking a $V_{SG1}=V_{SG2}$ diagonal line in Figure 2a would corresponds to a horizontal line going through the middle of the $\nu=2$ diamond in Figure 1c.) The prominent feature of Figure 2a is a square central region with resistance quantized at $R=h/2e^2$. When either side gate is applied beyond the plateau region, the resistance drops to a new quantized value. 

The observed influence of the side gates on the quantum Hall conductances is similar to the impact of local out-of-plane gates \cite{Tovari,Ozyilmaz}. The fact that the features in Figure 2a are purely horizontal or vertical shows that the influence of the two side gates is highly local: the left gate has a negligible effect on the right edge, and vice versa. This negligible cross-talk is different from that typically found in samples with out-of-plane gates. Furthermore, the side gates are very efficient, allowing us to control the filling factor of either edge in a wide range.

Figure 2b shows that the measured resistance drops from $R=h/2e^2$ to $R=h/6e^2$ if a positive side gate voltage is applied (green curve, measured along the green line in Figure 2a). This corresponds to $\nu_2$, the local filling factor on the side close to side gate 2 (SG2), reaching $\nu_2=6$ as shown schematically in Figure 2c. The bulk filling factor remains at $\nu=2$ and an additional conductance of $4e^2/h$ is contributed by the additional four-fold degenerate edge states induced near SG2. Note that in this case, the spatial separation between counterpropagating QH states in the side-gated region is less than 100 nm, as detailed further in the text.  The observation of quantized resistance plateaus suggests that backscattering between these counter-propagating states is suppressed, despite their close proximity. However, this should not be surprising, given that robust quantum Hall plateaus were previously observed in graphene nanoribbons of comparable width \cite{Morpurgo2012, Molenkamp2012}.

Next, the red line of Figure 2b demonstrates that each side gate can induce an independent $\nu=6$ state along its edge. Here SG1 is fixed at 3 V; this corresponds to a local filling factor near SG1 of $\nu_1=6$. Before SG2 is applied we start with resistance of $h/6e^2$: the baseline conductance is $2e^2/h$ and the right edge contributes additional $4e^2/h$, much like at the end point of the green curve in Figure 2b. Applying SG2 then adds an additional four-fold-degenerate channel on the other edge of the sample, resulting in the drop of resistance to $h/10e^2$, which corresponds to conductance of $(2+4+4)e^2/h$. 

Finally, we tune the back gate to 1.5 V (instead of 4.7 V), resulting in a bulk filling of $\nu=-2$. Applying SG1 now yields a transition from $R=h/2e^2$ to $R=h/4e^2$ (blue curve in Figure 2b.) The schematics in Figure 2d shows that in this case the side gate locally induces a QH state with an opposite filling factor of $\nu=2$, and the resulting plateau has a conductance of $(2+2)e^2/h$. Note that here as well, counterpropagating states are created in close proximity to each other. 

\begin{figure}[]
		\center \label{fig3}
		\includegraphics[width=3.1in]{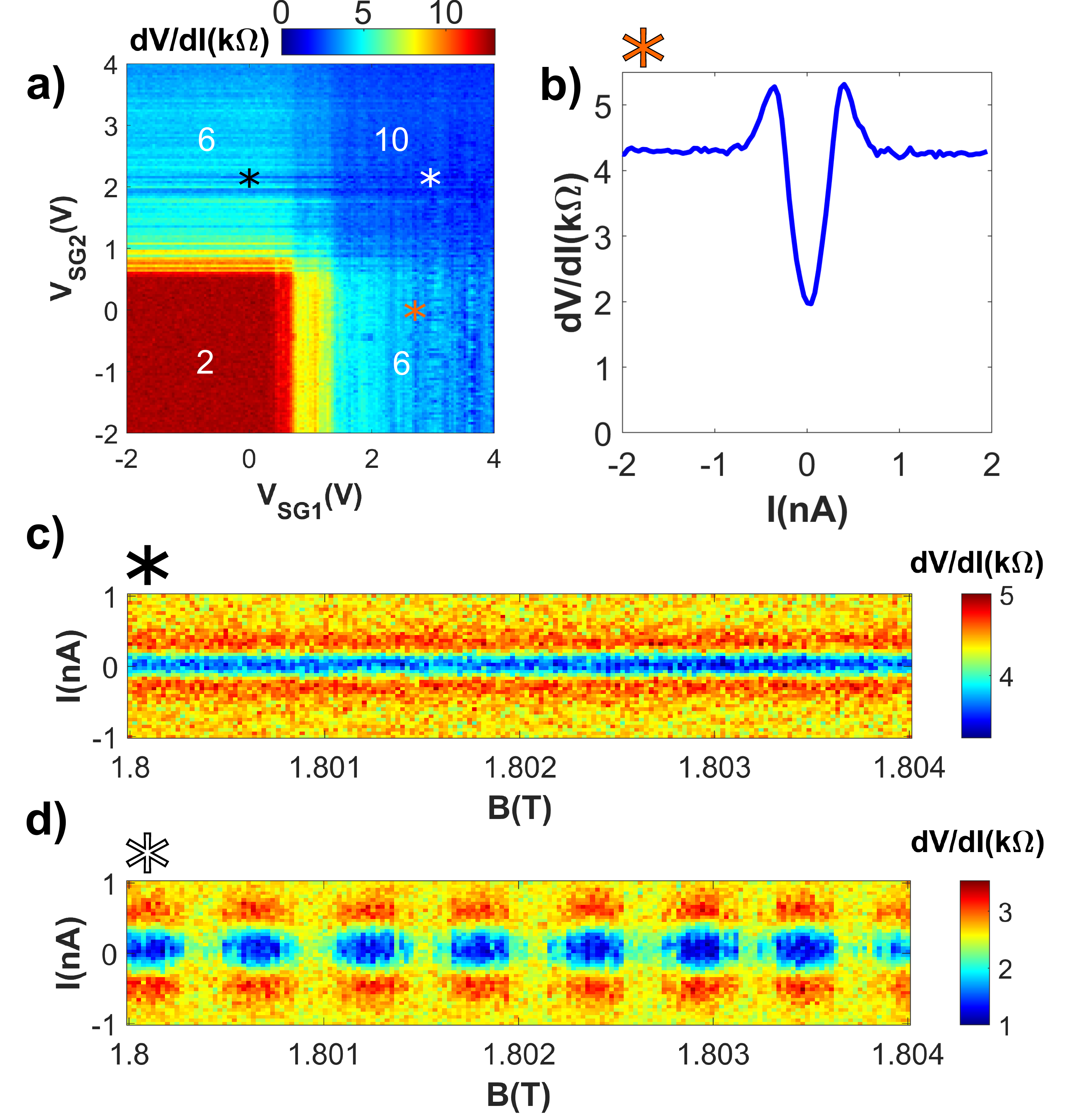}
		\caption{Quantum Hall supercurrent and its interference patterns. a) Differential resistance map vs. $V_{SG1,2}$ as Figure 2(a) but measured with 0 nA DC current bias, allowing observation of suppressed resistance due to the supercurrent. The gate voltage locations of panels (b-d) are marked by (b) an orange asterisk, (c) a black asterisk, and (d) a white asterisk. b) $dV/dI$ measured vs. $I$ indicating the presence of supercurrent on top of the quantized $h/6e^2$ plateau. c) Current - magnetic field map of the differential resistance when supercurrent is induced  along one side of the sample only with $V_{SG2}$, while $V_{SG1}$ stays at zero. The supercurrent is not sensitive to an incremental change of field on a few mT scale. d) A similar map with both side gates inducing supercurrent, showing a SQUID-like interference pattern.} 
	\end{figure}

\section{Side gates and quantum Hall supercurrent}

So far, the measurements have been performed with an applied DC bias current $I$ of 10 nA to suppress any supercurrent. We now switch $I$ to zero and explore the emerging superconducting features, maintaining the small AC current of 50 pA used to measure the differential resistance. Figure 3a shows a map of sample resistance vs. side gates similar to that in Figure 2a. While no supercurrent is found on top of the $\nu=2$ plateau, once the $\nu=6$ state is induced by either side gate the sample resistance develops pronounced dips that were not present at high DC current.

Figure 3b shows the sample resistance vs. bias taken at the location in Figure 3a marked by an orange asterisk, corresponding to $V_{SG2}=0$ V and $V_{SG1}=2.5$ V, so that $\nu_2$ is close to bulk filling and $\nu_1=6$. The region of suppressed resistance flanked by peaks is characteristic of a small supercurrent washed by thermal fluctuations. Notice that when the density enhancement is induced on one side only (regions in Figure 3a corresponding to the normal resistance of $h/6e^2$) the supercurrent features appear as horizontal/vertical lines - they depend on one side gate and do not vary with the other side gate. This confirms that the supercurrent is localized at one side of the junction.

Furthermore, the supercurrent does not vary for small changes in magnetic field (Figure 3c), indicating that the area it encompasses does not enclose additional flux quanta for a few mT change in field. This observation limits the distance between the counterpropagating edge channels responsible for the supercurrent to no more than $\sim 100$ nm (see also Supplementary Figure S1c). This distance is comparable to the coherence length of MoRe, which facilitates the coupling of the edge states to the superconductor and explains the appearance of supercurrent when a side-gate is turned on.

The dependence of the supercurrent on magnetic field completely changes when both side gates are applied, creating supercurrents along the two edges of the sample. Figure 3d shows a map similar to Figure 3c but with both side gates applied ($V_{SG1}=3.04$ V, $V_{SG2}=2.11$ V, marked by a white asterisk in Figure 3a). The map demonstrates a SQUID-like interference pattern with a period of 0.6 mT, close to that of the low-field Fraunhofer pattern of this junction (0.7 mT). 

\begin{figure}[]
		\center 
		\includegraphics[width=3.5in]{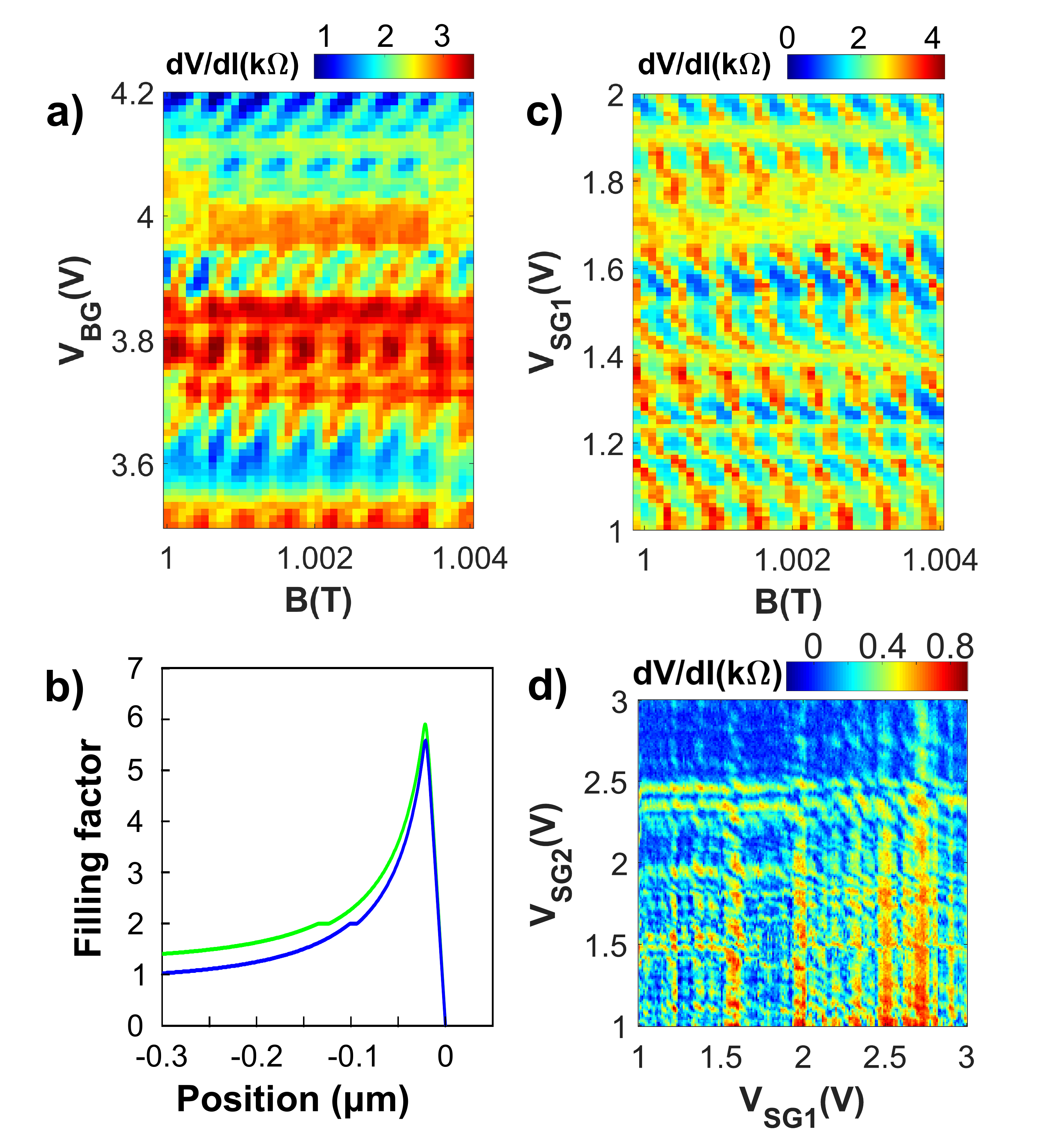}
		\caption{Side gate influence on supercurrent interference patterns. a) $dV/dI$ map measured at $V_{SG1}=2.34$ V and $V_{SG2}=2.36$ V as a function of $V_{BG}$ and $B$ near 1 T. For a given gate voltage, the regions of suppressed resistance correspond to stronger supercurrent. The pattern is periodic in $B$ with the same period as in Figure 3d. The phase of the oscillations depends on the gate voltage, indicating that the interference area decreases with the gate voltage (positive $dV_{BG}/dB$). This is explained by the inner edge states moving further inward as the electron density grows (schematic in b). b) Schematic of carrier density in the sample along the mid-line between the contacts. The blue line represents some baseline charge density; the green line shows a higher back gate voltage. These curves are generated using the electrostatic model discussed in the supplementary, but here are meant to be qualitative. c) $dV/dI$ map similar to (a) measured as a function of $B$ and SG1 voltage. The map shows an interference pattern with a slope opposite that in (a), indicating that the interference area increases with gate voltage as the electrons are pushed further toward the gate. d) Zero-bias $dV/dI$ map similar to Figure 3a, measured at 1 T and in a narrower $V_{SG}$ range.  Both side gate voltages are high enough to induce supercurrent ($V_{SG1,2} > 1$ V) and the vertical and horizontal features correspond to supercurrent induced by SG1 or SG2, respectively. At their intersections, additional diagonal features appear, indicating interference between the supercurrents on the two sides of the sample. The fringes have a slope $\sim -1$, suggesting comparable efficiency of the two side gates. }
		\label{fig4}
\end{figure}

We explore the supercurrent interference pattern in the QH regime as a function of gate voltages in Figure 4. Here, we change the field to 1 T in order to observe a more robust superconducting signature. Figure 4a shows the pattern of resistance oscillations in magnetic field, measured at zero applied DC bias as a function of the back gate. The period of the oscillations is found to be the same as in Figure 3d and independent of the gate voltage. The phase of the oscillations, however, is seen to vary with gate with an approximate slope of $+150V_{BG}$/T.

This gradual shift of the magnetic interference pattern with the back gate is explained by the fact that the changing electron density shifts the position of the QH edge states, thereby changing the area between the supercurrents on the two sides. Interestingly, the phase change from an increase of density (at more positive $V_{BG}$) is compensated for by the increase of the magnetic field, indicating that the effective area of the SQUID shrinks. This behavior can be understood from the schematic in Figure 4b, where the blue curve (lower back gate voltage) is compared to the green line (higher back gate voltage). The counterpropagating edge states occur on the opposite slopes of the non-monotonic density profile close to each edge. As the overall density increases (from the blue to the green curve), the inner states move further inward, while the outer states stay relatively stationary due to the very high density gradient close to the sample edge. As a result, on average the location of the supercurrent moves inward with increasing density.

A similar change in the interference pattern is observed when a side gate is applied (Figure 4c). The slope of this pattern is roughly $-300V_{SG1}/T$. Notably, the sign of the slope in Figures 4c is flipped compared to the one seen in Figure 4a. Following the discussion in the previous paragraph, this slope suggests that applying the side gate may be increasing the effective area of the SQUID. This could likely be attributed to the outward shift of the outer edge state, which is more strongly influenced by the side gate than the inner edge state. The very small size of the graphene region affected by the side gate might also result in charging effects, which are known to invert the slope of fringes  in quantum Hall interferometers \cite{Marcus2009, Ofek2010,Halperin2011}.

Finally, an additional interference pattern is revealed in Figure 4d, which shows the sample resistance as a function of both side gates ($B=1$ T). The interference is visible at the intersections of the vertical and horizontal lines corresponding to supercurrents flowing along the SG1 and SG2 edges, respectively. The interpretation of this interference pattern is similar to the discussion above, with each gate affecting the location of the edge state on its  side of the device. The contours of constant phase at the intersections of the vertical and horizontal lines have a roughly diagonal slope, indicating that the two gates have comparable efficiency. 

\section{Discussion}

We have shown that native graphene side gates are remarkably efficient in controlling edge state propagation in the quantum Hall regime. They enable full control of the local filling factors along the sample edges, allowing us to fill the next Landau level, change carrier polarity, or keep the density flat close to the edge. Further, we have observed supercurrents carried by the QH edge states induced by the side gates. These supercurrents flow independently on each edge of the device and could be controlled independently by the corresponding gates. Our experiment opens a promising route for coupling superconductors with QH edge states for the purpose of inducing non-Abelian excitations. 

\section{Methods}

The sample was made with mechanically exfoliated flakes of graphene and hexagonal boron nitride. It was assembled using a standard stamping technique \cite{Wang}. The resulting heterostructure was patterned using electron beam lithography followed by reactive ion etching with $CHF_3$ and $O_2$ to expose the edges of the encapsulated graphene. These edges were contacted with $100$ nm of molybdenum rhenium (50-50 ratio by weight) sputtered onto the etched regions. The device boundaries and sidegates were defined with a second round of lithography and etching.

Measurements were performed in a Leiden Cryogenics dilution refrigerator at a temperature of $\sim 100$ mK. The sample was electronically isolated in the refrigerator via resistive coax lines and low-temperature RC filters. Differential resistance measurements were carried out using an AC excitation current of 50pA. Magnetic fields for quantum Hall measurements were applied perpendicular to the sample plane. 

\section{Data availability}

The datasets supporting the figures and conclusions of the current study are available from the corresponding author upon request.

\section{Acknowledgments}

We thank Harold Baranger, Ady Stern, and Enrico Rossi for helpful discussions. Transport measurements conducted by A.S., A.W.D., and E.G.A. were supported by Division of Materials Sciences and Engineering, Office of Basic Energy Sciences, U.S. Department of Energy, under Award No. DE- SC0002765. A.S. and M.T.W. performed lithographic fabrication and characterization of the samples with the support of NSF awards ECCS-1610213 and DMR-1743907. G.F. was supported under ARO Award W911NF16-1-0122. H.L. and F.A. acknowledge the ARO under Award W911NF-16-1-0132. K.W. and T.T. acknowledge support from JSPS KAKENHI Grant Number JP15K21722 and the Elemental Strategy Initiative conducted by the MEXT, Japan. T.T. acknowledges support from JSPS Grant-in-Aid for Scientific Research A (No. 26248061) and JSPS Innovative Areas “Nano Informatics” (No. 25106006). This work was performed in part at the Duke University Shared Materials Instrumentation Facility (SMIF), a member of the North Carolina Research Triangle Nanotechnology Network (RTNN), which is supported by the National Science Foundation (Grant ECCS-1542015) as part of the National Nanotechnology Coordinated Infrastructure (NNCI).

\section{Author Contributions}

A.S., F.A., and G.F. designed the project and experiments. A.S., A.W.D., and E.G.A. conducted measurements and analyzed the data. G.F. supervised the experiments. K.W. and T.T. provided the crystals of hexagonal boron nitride. H.L. and F.A. assembled the graphene / boron nitride heterostructure. A.S. and M.T.W. fabricated the device. A.S., F.A., and G.F. wrote the manuscript.

\section{Competing financial interests}
The authors declare no competing financial interests.

\clearpage

\graphicspath{{/}{Supplementary/}}

\pagebreak
\onecolumngrid
\begin{center}
\textbf{\large Supplementary material: Full control of quantum Hall supercurrent in a side gated graphene Josephson junction}
\end{center}

\setcounter{figure}{0}
\renewcommand{\thefigure}{S\arabic{figure}}
\setcounter{section}{0}

   In support of the data presented the main paper, here we include additional measurements and types of differential resistance maps.

\section{Additional supercurrent interference maps}

Figure S1 expands on the 1.8 T data presented in Figures 1-3 in the main text. Panel a shows a wider version of the SG1-SG2 map of differential resistance seen in Figure 3a with $I_{DC}=0$ nA. Here, in addition to the side gate induced quantum Hall plateaus, we see lines of reduced resistance as supercurrent appears at certain side gate voltages on either side of the junction. 
The panels of Figure S1b show the magnetic interference pattern of one supercurrent pocket induced by SG2 as SG1 changes from 0 to +3 V, following the dashed line in Figure S1a.
Once both gates are $\geq 2$ V, the supercurrents along the two edges interfere and we see the transition from an aperiodic to a fully SQUID-like interference pattern.
Finally, Figure S1c shows aperiodic resistance measured as a function of magnetic field and SG1, with SG2 held at 0 V. This contrasts with the periodic side gate - field interference map taken in the regime when both edges carry supercurrent (Figure 4c). Clearly, since the supercurrent is independently controlled along each edge, only when both edges are active is there interference. As stated in the main text, we interpret the interference observed in Figure 4d as the result of each side gate tuning the location of its associated edge states, thus altering the enclosed magnetic flux through the junction. 

Figure S2 presents additional interference data at $B=1$ T. 
S2a and b contrast the differential resistance maps taken at $I_{DC}=10$ nA and $I_{DC}=0 nA$; S2b is an extended version of the map in Figure 4d. It is clear that the map in S2a lacks the diagonal interference features present in S2b. Indeed, these features are attributed to interference of supercurrents, which are suppressed by $I_{DC}=10$ nA. Figures S2c,d detail the transition of the side gate interference pattern from periodic to aperiodic as a function of each side gate. This is consistent with the picture of independent supercurrents on each edge of the device being present at only certain local gate voltages.

\section{Measurements around the bulk $\nu=6$ plateau}

The main text focuses on data taken when the bulk of the sample is tuned to the $\nu=2$ plateau. In this section, we include complementary data taken at the $\nu=6$ plateau in the bulk, at $B=1$ T.
In Figure S3a, increasing either side gate leads to the total differential resistance reaching $h/10e^2$. This indicates the addition of a fourfold degenerate state only along one side of the junction. Tuning the second gate, understandably, yields $h/14e^2$: Each side of the junction can be thought of as locally being in the $\nu=10$ state, while the bulk remains at $\nu=6$. Each edge contributes $4e^2/h$ in addition to the base $6e^2/h$ conductance.
Figure S3b presents a map of differential resistance at $I_{DC}=0$nA, showing a multitude of locations with superconducting pockets. 

\section{Measurements of a second device}

In this section we present measurements similar to those in the main paper, but taken on a different device (J2). The dimensions of J2 are similar to J1, and the junction region is also separated from the two side gates by  $\sim 60$ nm - wide trenches. However, here the contacts are not spaced from the trenches by 100 nm regions of graphene, as was done in J1. This second device was fabricated on the same chip as the main device. In fact, SG2 is shared by both junctions (Figure S4a).

The data presented were taken at $B= 1$ T with $V_{BG}=1.85$ V in the center of the $\nu=-2$ plateau. Side gate - side gate maps of differential resistance are presented in Figure S4b,c and show the development of new resistance plateaus at high side gate voltage. These correspond to $h/4e^2$ when one side gate is applied, and to $h/6e^2$ when both are applied. These values are explained by the appearance of $\nu=2$ channels along each side of the junction. (See Figure 2d in the main text.)

Like our main junction J1, this device shows regions of supercurrent in the $I_{DC}=0$ nA bias map (Figure S4c). As seen previously in Figure S1d, the dependence of supercurrent on magnetic field also undergoes a transition from non-periodic to SQUID-like: compare Figure S4d, in which only SG2 gate is active, to Figure S4e, in which SG3 is also applied, resulting in interference. 

\section{Electrostatic Simulations}

In order to simulate the carrier density profile induced by the side gate in the quantum Hall regime, we determined the geometric local capacitance between the graphene sheet and both the back gate and side gates. 

We solve the Laplace equation for the electrostatic potential with Dirichlet boundary conditions at the back gate, side gates, and the graphene sheet. Both local capacitances are spatially varying and are stronger at the graphene edge as a result of electric field focusing.

The back gate and the intrinsic doping $n_{0}$ of the graphene sheet (determined experimentally by the location of the Dirac peak in a gate sweep of the device) were then used to determine the bulk carrier density in the graphene and subsequently the chemical potential. 

Applying a positive voltage on the side gate tends to raise the graphene potential near the edge. At zero temperature the local graphene potential $V_{g}(x)$ can be obtained by solving: 

\begin{equation}
n_{0}+C_{BG}(x)(V_{BG}-V_{g}(x))+C_{SG}(x)(V_{SG}-V_{g}(x)) = \int_{0}^{\mu+V_{g}}\rho(E)dE.
\end{equation}
Here, we defined the density of states as a sum of Gaussians centered at the Landau level energies $E_{n}$ with a Landau level degeneracy $N$:

\begin{equation}
\rho(E)=N \sum_{n=0}^{\infty} \frac{1}{\tau\sqrt{2\pi}}exp( -\frac{1}{2}(\frac{E-E_{n}}{\tau})^{2})
\end{equation}
where $\tau$ parametrizes the breadth of the Landau levels. The local carrier density is then:

\begin{equation}
n(x)=\int_{0}^{\mu+V_{g}}\rho(E)dE.
\end{equation}

The vanishing of the carrier density at the edge was artificially obtained by linearly bringing the density to zero at the edge over a length scale l. The length scale of the convergence was chosen to be on the order of the magnetic length ($\sim 20$nm) and justified post hoc through comparison of the simulated and experimental gate maps. This behavior is of course a simplification and further theoretical work will be needed to understand the evolution of the carrier density in the few nanometers near the edge.

Figure S5 details the evolution of carrier density within 300nm of the junction edge with the application of back gate (Figure S5a) and side gate (Figure S5b) voltages. Figure S5a quantitatively builds on Figure 4b from the main text. We note that the electrostatic simulation of the side gate influence in Figure S5b does not reproduce the movement of the supercurrent carrying states towards the edge of the junction, as is discussed in the main text in explaining the slope of the interference pattern shown in Figure 4c.

Finally, to reproduce the differential resistance map shown in Figure 1c, the carrier density was used to determine the number of edge states in the system and plotted as a function of gate, generating Figure 1d. Commensurate with the integer plateaus seen in the experiment, we considered only the degenerate graphene filling factors ($\nu=2,6,10,\dots$) and added in the emerging $\nu=0$ peak (defined in the model for densities corresponding to $n_{\nu=-1}<n<n_{\nu=1}$), which was assigned an arbitrary conductance for map contrast (dark red in the map of Figure 1d).

\onecolumngrid

\begin{figure}[]
		\center \label{figS1}
		\includegraphics[width=7in]{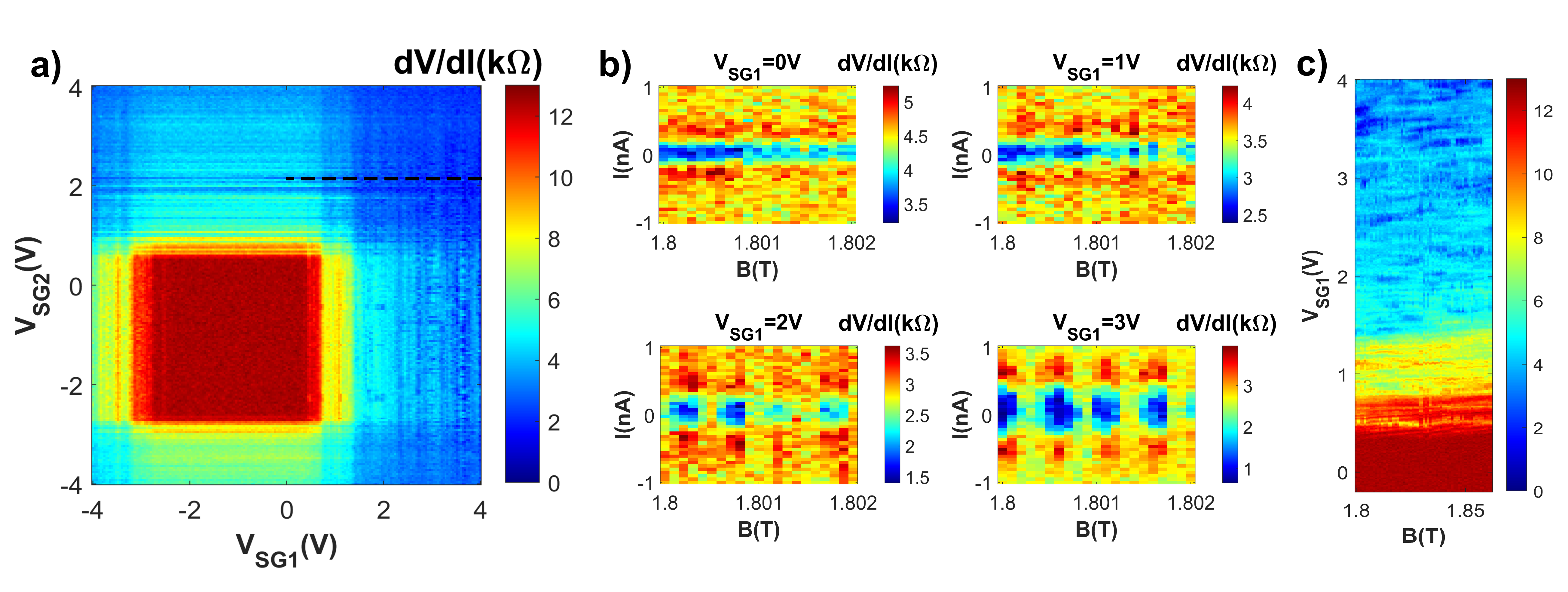}
		\caption{Additional side gate maps and interference patterns at 1.8 T. a)  Differential resistance map measured as a function of both side gates at $I_{DC}=0$ nA and bulk filling $\nu=2$. This is an extended version of the map shown in Figure 3a of the main text.  Large enough negative side gate voltages result in transitions from the central $\nu=2$ plateau to the regions of differential resistance equal to $h/4e^2$ (as explained in Figure 2d). Supercurrent pockets again appear at certain side gate values as vertical or horizontal lines of suppressed resistance. 
		b) Evolution of the supercurrent vs. $B$ maps as SG1 grows from 0 to 3 V along the dashed line in (a). The maps show the transition from aperiodic supercurrent at $V_{SG1}=0$ V to a SQUID-like periodicity at $V_{SG1}=3$ V. This illustrates the change of behavior from supercurrent on one side of the junction to supercurrent on both sides.
		c) Map of differential resistance as a function of SG1 and magnetic field, taken at 0 DC bias. This is similar to Figure 4c, but here $V_{SG2}=0$ V, so supercurrent is induced only along SG1. The supercurrent is clearly aperiodic, showing only some irregular dependence on magnetic field. Note that compared to Figure 4a,c of the main text, the range of fields here is about 10 times wider. }
	\end{figure}
	
	\begin{figure}[]
		\center \label{fig2}
		\includegraphics[width=4in]{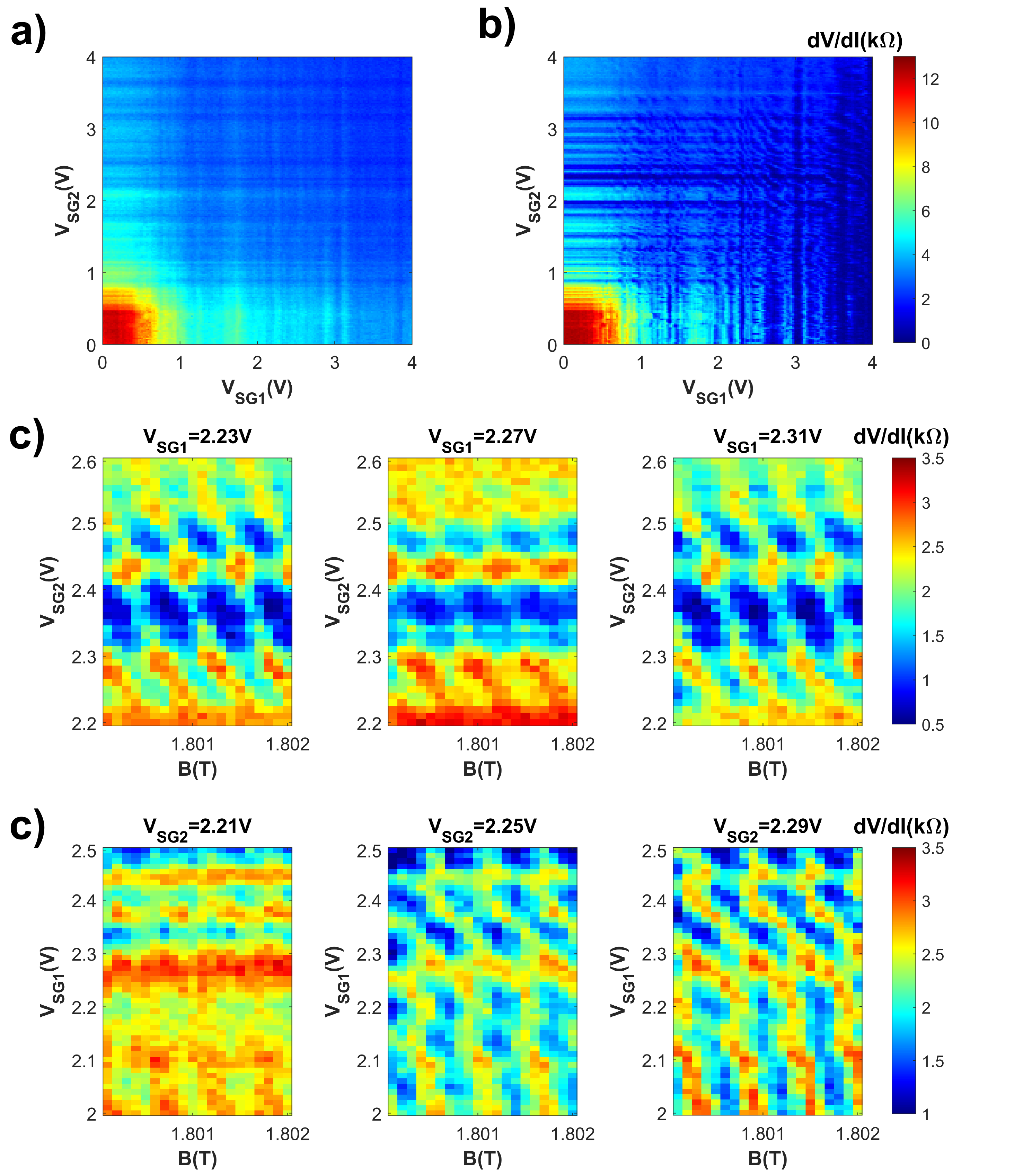}
		\caption{Additional side gate maps and interference patterns at 1 T. a-b) Wider SG1-SG2 maps of differential resistance at a) 10 nA and b) 0 nA DC bias. The map in (a) lacks the diagonal interference features of (b), which are interpreted as a supercurrent interference effect and so are not present in the 10 nA bias condition.
		c) Maps of differential resistance measured as a function of SG2 and field $B$ at 0 DC bias. The maps are taken at three SG1 locations (2.23, 2.27, 2.31 V) in the regime when the two supercurrents interfere. The observed patterns are very similar, except for the shift of the phase of the oscillations and the overall contrast. These changes are explained by the change in the strength and geometrical position of the supercurrent flowing along along SG1 as the corresponding voltage is changed between the three maps. Note that at the bottom of all maps ($V_{SG2}=2.2$ V) the pattern becomes roughly independent of magnetic field, because at that value of SG2 the corresponding current is equal to zero. d) Similar to (c), but here SG1 is swept and SG2 is fixed at 2.21, 2.25, and 2.29 V. Note that the left map taken at $V_{SG2}=2.21$ V is barely periodic because the current flowing along SG2 is very small.}
	\end{figure}

\begin{figure}[]
		\center \label{fig3}
		\includegraphics[width=4.5in]{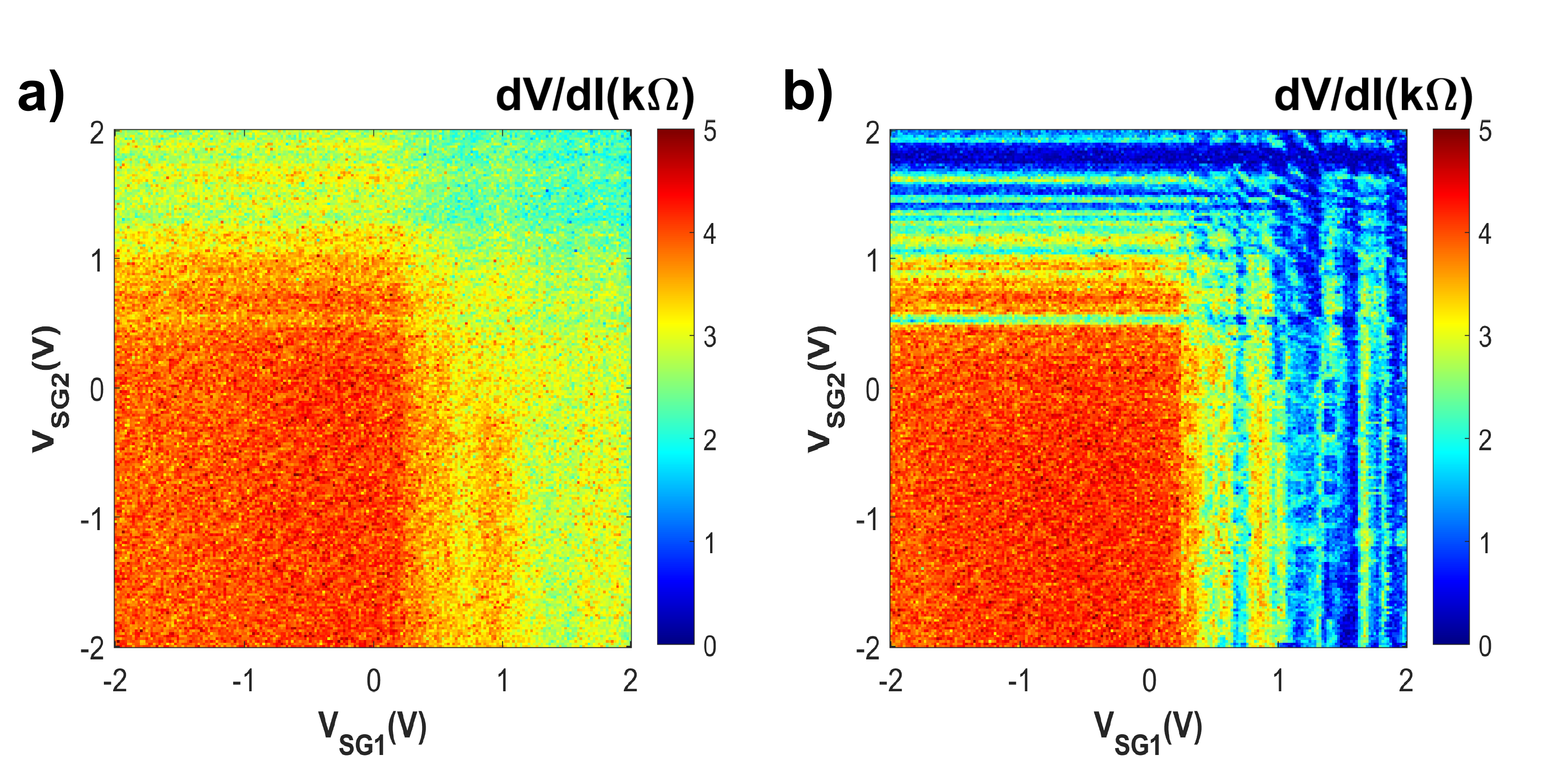}
		\caption{Supercurrent at $\nu=6$ in the bulk at 1 T. a-b) SG1-SG2 maps of differential resistance taken at back gate voltage $V_{BG} = 5.55$ V, which corresponds to $\nu=6$ filling in the bulk. 
	The maps are measured at a) 10nA and b) zero DC bias. In a) the plateaus have resistances of $h/6e^2$ (side gate equal to zero or negative), $h/10e^2$ (one side gate applied), and $h/14e^2$ (both side gates on). The SG1-SG2 interference effect shown in Figure 4e is also seen here in panel (b).  }
	\end{figure}

\begin{figure}[]
		\center \label{fig4}
		\includegraphics[width=7in]{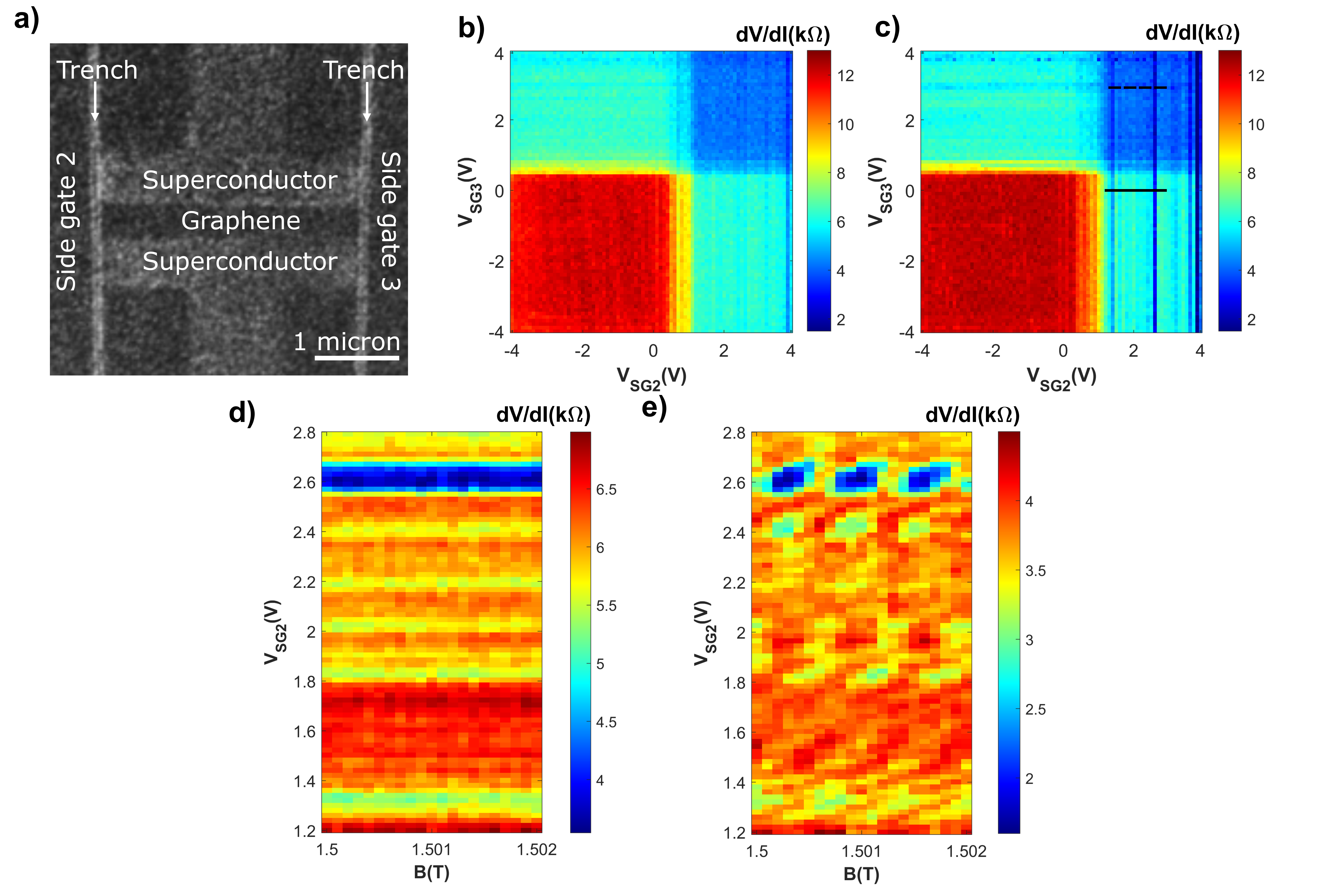}
		\caption{Study of a second device at 1 T. a)  SEM micrograph of the device prior to reactive ion etching. MoRe contacts are again visible in lighter gray. Two trenches, $\sim 60$ nm wide, separate the junction from the side gates. The side gates are labeled SG2 and SG3, where SG2 is shared with the device from the main paper. b-c) SG2-SG3 maps of differential resistance taken at b) 10nA and c) 0na DC bias like in the previous figures. The maps are taken at $V_{BG}=1.85$ V, corresponding to $\nu=-2$. Panel (b) shows that one side gate induces a resistance of $h/4e^2$, and both side gates induce a resistance of $h/6e^2$, as expected from the study of the first device. Panel (c) shows the development of supercurrent at positive SG voltages.  The solid and dotted lines show the location of the maps in (d) and (e), respectively. d) Map of differential resistance as a function of SG2 and field at $V_{SG3}=0$ V. Aperiodic supercurrent is seen (e.g. see the pockets at $V_{SG2}=$1.3 and 2.6 V), which again indicates the localization of the supercurrent along one edge. e) Similar map of SG2 vs. field at $V_{SG3}=2.9$ V showing a periodic supercurrent pattern that varies with gate. The sloped features show that the phase of the oscillations depends on gate voltage, indicating that the interference area changes with gate voltage, similar to the result of Figure 4c of the main text.}
	\end{figure}
	
	\begin{figure}[]
		\center \label{fig5}
		\includegraphics[width=5in]{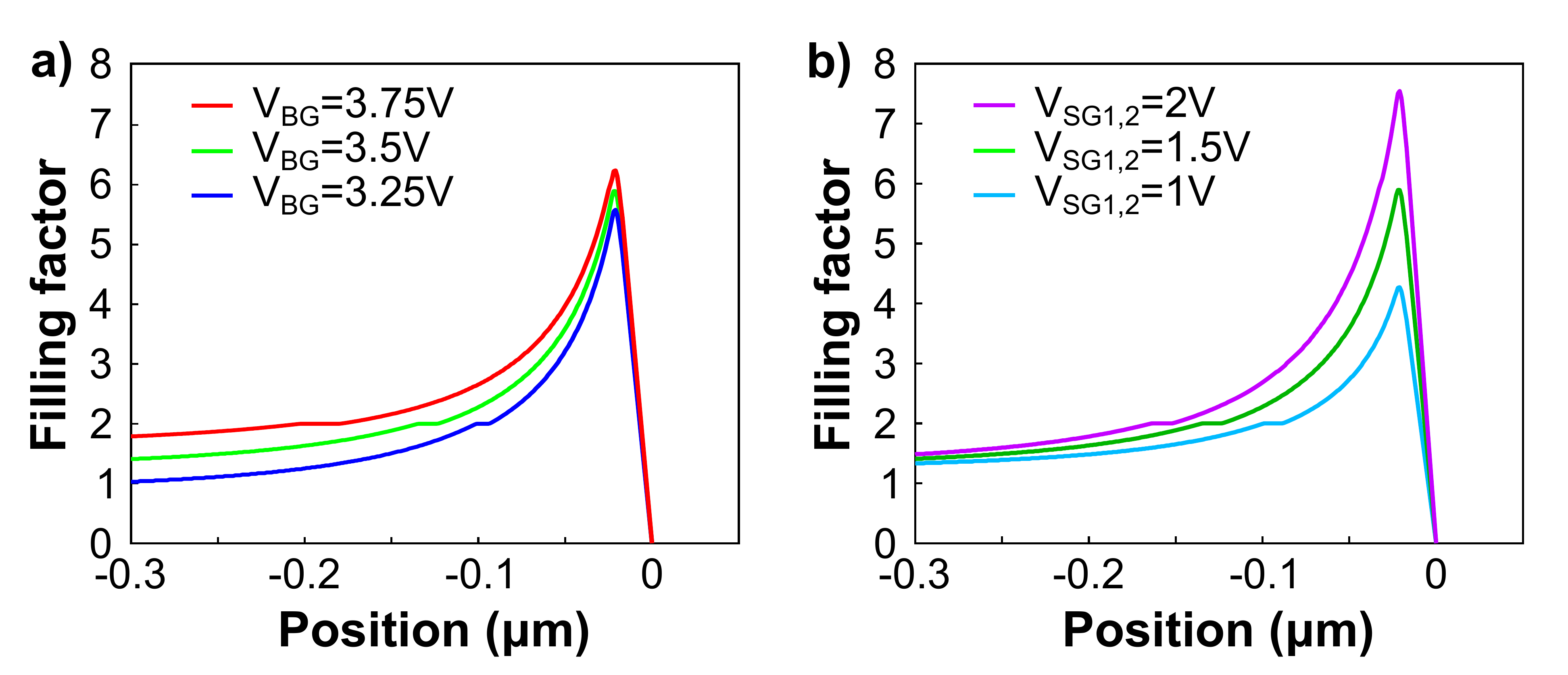}
		\caption{Simulated evolution of carrier density near the junction edge. a) Simulated carrier density at several back gate locations (blue 3.25 V,  green 3.5 V, red 3.75 V) within 300 nm of the sample edge with both side gates held at 1.5 V. This is as Figure 4c of the main text with the addition of the red curve and gate values given for quantitative comparison. Increasing the gate voltage is seen to impact the edge density somewhat and to have a strong impact on the bulk density away from the edge. The flat features are incompressible strips resulting from the quantum Hall effect. b) As (a) with $V_{BG}=3.5$ V, but with the side gate location tuned (cyan 1 V, green 1.5 V, purple 2 V). The green curves on both (a) and (b) are identical. The side gate voltage is seen to significantly shift the edge density with only a modest impact even 300 nm into the bulk.}
	\end{figure}

\end{document}